# Demonstration of atom interrogation using photonic integrated circuits anodically bonded to ultra-high vacuum envelopes for epoxy-free scalable quantum sensors

STERLING E. MCBRIDE[1,*], CALE M. GENTRY[2], CHRISTOPHER HOLLAND[3], COLBY BELLEW[3], KAITLIN R. MOORE[4], AND ALAN BRAUN[1]

[1]*Advanced Technology and Systems Division, SRI International, 201 Washington Road, Princeton, NJ 08540, USA*
[2]*Advanced Technology and Systems Division, SRI International, 2595 Canyon Blvd, Suite 440, Boulder, CO 80302, USA*
[3]*Advanced Technology and Systems Division, SRI International, 333 Ravenswood Ave, Menlo Park, CA 94025, USA*
[4]*Advanced Technology and Systems Division, SRI International, 2100 Commonwealth Boulevard Third Floor, Ann Harbor, MI 48105, USA*
*\*sterling.mcbride@sri.com*

**Abstract:** Reliable integration of photonic integrated circuits (PICs) into quantum sensors has the potential to drastically reduce sensor size, ease manufacturing scalability, and improve performance in applications where the sensor is subject to high accelerations, vibrations, and temperature changes. In a traditional quantum sensor assembly, free-space optics are subject to pointing inaccuracies and temperature-dependent misalignment. Moreover, the use of epoxy or sealants for affixing either free-space optics or PICs within a sensor vacuum envelope leads to sensor vacuum degradation and is difficult to scale. In this paper, we describe the hermetic integration of a PIC with a vacuum envelope via anodic bonding. We demonstrate utility of this assembly with two proof-of-concept atom-interrogation experiments: (1) spectroscopy of a cold-atom sample using a grating-emitted probe; (2) spectroscopy of alkali atoms using an evanescent field from an exposed ridge waveguide. This work shows a key process step on a path to quantum sensor manufacturing scalability.

## 1. Introduction

There is a growing interest in the miniaturization of quantum sensors by directly integrating neutral-atom or ion platforms with photonic integrated circuits (PICs) [1-6]. Applications include the miniaturization of cold-atom sensors [7], the development of dense arrays of fieldable quantum sensors [3,8] and atom-based frequency references [2,6]. In most atom-based, state-of-the-art quantum sensors, laser light is delivered using free-space beams across an air-vacuum interface via optical windows. This approach requires discrete optical elements such as mirrors, lenses, polarization optics, and beamsplitters. Recently, various groups have demonstrated such optics functions on photonic chips for beam routing and delivery [9], demonstrating photonics configurations suitable for atom cooling [10-12], atomic magnetometry [3], and the interaction of alkali vapors with micro-ring resonator structures [2,6]. In these demonstrations, until now to the best of our knowledge a PIC is integrated with an alkali-containing vacuum or buffer-gas enclosure either via a glass interposer or epoxy bonding.

This paper presents a photonic design and epoxy-free, anodic-bonding process to integrate a silicon-dioxide ($SiO_2$)-clad silicon-nitride (SiN) PIC with an ultra-high vacuum (UHV) enclosure, in which the PIC forms a structural part of the enclosure wall. The utility of this integration is demonstrated via two proof-of-concept atom-interrogation experiments. While the demonstrations in this paper show bonding of a PIC to the edges of a glass cuvette, this process is readily extendable to a bulk silicon envelope. Such a development could lead to the architecture of a fully enclosed UHV envelope to support light-atom interactions without bulk glass, mitigating helium permeation and eliminating free-space optics to produce devices robust against mis-aligning thermal and vibration effects.

The main challenges solved by the design and process presented in this paper are (1) materials compatibility, e.g., using materials suitable for anodic bonding; (2) maintaining the integrity of the passive photonic structures on the PIC (e.g., thermal compatibility with UHV processing and bonding temperatures); (3) utilizing less stringent surface flatness requirements of the PIC than for techniques like fusion bonding; and (4) forming vacuum-tight seals without glues or epoxies. Glues and epoxies are not a long-term solution for quantum sensors because they contaminate vacuum, especially for small-size (few cubic-centimeter) UHV cell volumes. The processes presented in this paper are fully compatible with established chip-scale quantum-sensing batch-fabrication technologies [1,3,7,13] and introduce the achievement of the important and challenging step of PIC integration with the vacuum envelope via anodic bonding.

## 2. Methods

The demonstrated solution involves the combination of two key developments: (1) the application of a glass-silicon stack to support leak-free, epoxy-free anodic bonding of the photonic chip as one wall of the vacuum chamber, and (2) the design and demonstration of a photonic "bridge" structure to deliver guided light through the air-vacuum hermetic seal with very low loss. Using these processes, we have custom-built six successful photonic-UHV systems over the past two-plus years. To demonstrate the utility of two of these builds, this paper describes two experiments: one experiment using a PIC-vacuum system to provide a grating-emitted free-space probe laser for interrogation of a magneto-optical trap (MOT) and another experiment in a second, identically fabricated system using a PIC-vacuum system to provide an evanescent-field probe for interrogation of rubidium vapor in a cold-atom vacuum system. In the following sub-sections, we describe the anodic bonding process, the design of a photonic structure for low light loss across the air-vacuum interface, the design and fabrication of the photonic grating emitter for the first experiment, and the design and fabrication of the photonic evanescent loop for the second experiment.

### 2.1 Epoxy-free vacuum seal using photonic chip

To demonstrate the integration of a PIC and vacuum, we first prepare a cuvette and UHV assembly. The assembly consists of an 18 mm × 18 mm outer cross-section glass cuvette attached to a standard glass-metal transition with a ConFlat flange, for assembly with a pump (Agilent VacIon, 2 L/s) and alkali dispenser (SAES RB/NF/3.4/12FT10). The bonding edges of the glass cuvette are polished to a flatness of $< \lambda/2$, as determined by use of an optical flat.

Next, an amorphous silicon (a-Si) layer is applied to the PIC $SiO_2$ layer to achieve bonding [14]. We note that in our experience, we have been able to bond without any further processing, such as chemical mechanical polishing (CMP).

Finally, the PIC is anodically bonded to the glass cuvette at temperatures below 300 ºC, forming one of the walls of the vacuum enclosure. We measure helium leak rates of $< 10^{-10}$ Torr L/s at the bonded interface, below the detection limit of the helium leak detector.

In Fig. 1 we show two examples of PICs integrated with vacuum enclosures. The first example, shown in Fig. 1(a), is a PIC with a grating close to the center of the chip. The stack-up in Fig. 1(a) includes: (1) PIC substrate with a-Si layer; (2) diced borosilicate (Pyrex) glass chip with micro-machined aperture to expose the grating out-coupler; (3) silicon square frame; (4) borosilicate (Pyrex) glass cuvette with polished edges. This is an early prototype. We include this stack-up to demonstrate that in future designs, the glass chip may be thinned, the glass cuvette eliminated, and the silicon frame replaced by a bulk silicon enclosure, to eliminate all bulk glass in this system.

The second example, shown in Fig. 1(b), demonstrates integration of a multi-function PIC using a simplified stack-up that includes: (1) PIC substrate with a-Si layer; (2) borosilicate (Pyrex) glass cuvette with polished edges. The multi-function PIC is designed to form an evanescent field optical trap (EFOT) using an unclad waveguide loop structure, a magnetic wire trap using a gold trace in proximity to the waveguide loop, and a mirror-MOT using a mirror metal coating over most of the surface of the PIC. The mirror metal coating is gold, and it is protected by a $Al_2O_3$ thin film. This PIC design is described in further detail in Sec 3.2. For the multi-function PIC, the gold trace is additionally fed through the bonded seal to the edge of the chip by using an interposer chip with UHV compatible electrical feedthroughs. The interposer chip is anodically bonded to the back of the PIC. This process creates a vacuum hermetic seal between the PIC and the glass cuvette, with a background vacuum pressure of $<10^{-10}$ torr with the rubidium dispenser turned off.

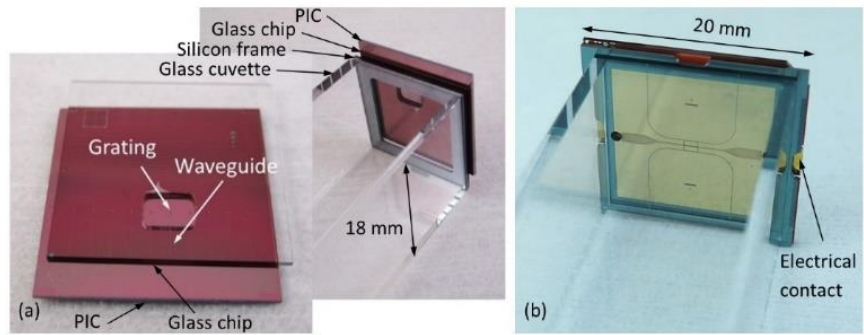

Fig. 1. Photonic integrated circuits anodically bonded to 18 mm × 18 mm glass cuvettes, forming a wall of the vacuum enclosure: (a) PIC with a grating emitter; (b) PIC with unclad, protected photonic circuit for evanescent interrogation, magnetic trapping wire, and mirror.

Finally, using the process described above, we have also demonstrated the anodic bonding to a PIC of a linear array of micromachined cells in a borosilicate (Pyrex) glass substrate, demonstrating the feasibility of fabricating vapor cells at the wafer level where one of the walls is a PIC for light delivery and detection. Fig. 2(a) shows a picture of a 4-cell array in a glass substrate that is anodically bonded to a PIC, forming a hermetic seal. To form the cells, 2.5 mm x 2.5 mm pockets were micromachined in a 3 mm thick borosilicate glass substrate. Fig. 2(b) shows a closeup picture of the cells, showing 2 gratings per cell that are individually routed by waveguides to the edge of the PIC for light coupling.

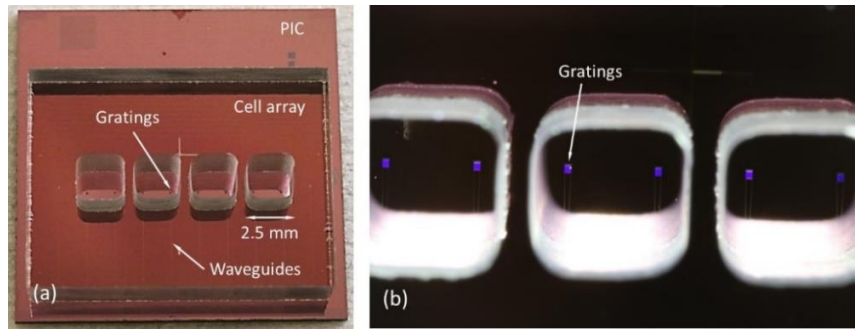

Fig. 2. Glass cell array anodically bonded to a linear array of gratings on a PIC: (a) linear array of 4-cells micromachined in a 3 mm thick glass substrate; (b) closeup view of the cell openings, showing 2 gratings per cell.

## 2.2 Photonic bridge structure

We describe a PIC waveguide design for high-efficiency optical guiding across the air-vacuum interface, while maintaining UHV hermeticity. Overall, the PIC consists of an edge-coupled $SiO_2$-clad SiN ridge waveguide that tapers from a single mode (400 nm wide by 200 nm thick) waveguide to a 5-micron wide waveguide over a length of 500 microns. In Fig. 3(a) we show the schematic of the 5-micron wide waveguide at the vacuum wall interface. Compared to a single-mode waveguide, this waveguide supports a lower electric field at the waveguide sidewalls, resulting in less scatter at the interface with the vacuum wall. In addition, the wider optical mode provides less diffraction when launched into the slab region. At the vacuum wall interface, the mode of the waveguide propagates without a sidewall for 7 microns in a $SiO_2$-clad SiN "bridge" structure, allowing for a fully continuous bonding region and hermetic seal. After the vacuum wall interface, the waveguide tapers back to single-mode width.

In Fig 3(b), we show simulation results of 780-nm light propagating through the vacuum wall interface. The efficiency of the wall structure was simulated using a three-dimensional (3-D) finite-difference time-domain (FDTD) model, which demonstrated losses of 2%. We find that efficiencies greater than 99% are possible with a vacuum wall less than five microns wide, which we have demonstrated in a separate experiment still supports UHV hermeticity.

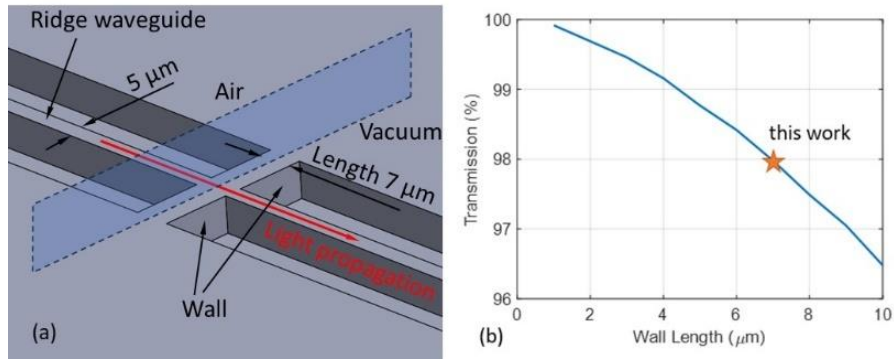

Fig. 3. Photonic bridge structure. (a) Schematic of $SiO_2$-clad SiN vacuum wall ("bridge") structure in $SiO_2$-clad SiN photonic ridge waveguide. At the vacuum wall interface, the mode of the waveguide propagates without a sidewall for 7 microns, allowing for a fully continuous bonding region and hermetic seal. (b) 3-D FDTD simulation of 780-nm light propagating (and diffracting) through the bridge structure, plotting simulated transmission efficiency vs. wall length (blue line) and indicating transmission efficiency possible for the 7-micron design used in this work (star).

## 2.3 Generation of grating-emitted free-space probe

For the first proof-of-concept experiment described in Sec. 3.1, we have designed and fabricated a PIC with a grating structure to emit a free-space beam inside the anodically-bonded vacuum enclosure. As described in Sec. 2.2, a ridge waveguide with a bridge structure guides light from the edge of the chip on the air side to the grating on the vacuum side. The hafnium oxide ($HfO_2$) grating has been fabricated with a pitch of 510 nm and tooth width of 301 nm (i.e., fill factor of 59%). The PIC for this demonstration was processed on a 100 mm silicon wafer. The wafer had a 4-micron thick thermal oxide and a 200-nm thick $Si_3N_4$ coating deposited by liquid-phase chemical vapor deposition. The waveguide features were patterned into photoresist using an I-line ASML stepper into Shipley SPR-955 resist, spun at 5000 rpm to yield a 0.7-micron thick film. The resist pattern was transferred into the $Si_3N_4$ with a $CHF_3:O_2$ (60:10.5 sccm) chemistry at 50 mTorr using a Plasma-Therm 7200 reactive-ion etching system.

After etching and resist stripping, the wafers were spin-coated with two layers of different molecular weight PMMA anisole-based resists: (1) 495K and (2) 950K. Both these layers were individually hot-plate-baked at 180 ºC for 2 minutes. The grating couplers were patterned into the PMMA using a JEOL JBX-6300FS e-beam lithography system and developed in a 1:3 solution of MIBK:IPA. The wafers were then coated by e-beam evaporation with 100 nm $HfO_2$. Then, the wafers were soaked in acetone releasing the $HfO_2$ everywhere except the gratings. The wafers were then coated by plasma-enhanced chemical vapor deposition (PECVD) with 1200 nm $SiO_2$ and subsequently diced into 20-mm square chips.

In Figs. 4(a-c) we show a diagram and simulations supporting the chosen grating emitter design. In Fig. 4(d) we show a photographic image of a grating overlapping the tapered $Si_3N_4$ waveguide, which is approximately 100 microns wide at the grating location and 400 nm wide at the edge of the PIC.

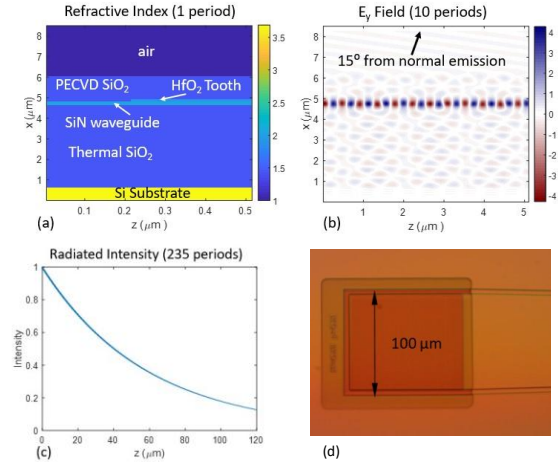

Fig. 4. (a) Diagram of refractive index (colorbar) cross-section of a single grating period. The stack starts on a silicon substrate followed by thermal oxide, silicon nitride waveguide, and $HfO_2$ grating, all covered by a PECVD silicon oxide thin film. (b) Simulated electric field (colorbar, a.u.) distribution over 10 grating periods demonstrating that a $HfO_2$ thickness of 100 nm provides out-of-plane grating emission of 15 degrees from normal. (c) Simulated grating strength is plotted as radiated intensity over the length of the grating, corresponding to a field decay factor of 8.3 1/mm. (d) Photographic image (top view) of fabricated 100 µm × 100 µm $HfO_2$ grating with a 59% fill factor on top of a tapered optical waveguide.

### 2.4 Generation of evanescent field probe

For a second proof-of-concept experiment described in Sec. 3.2, we have designed and fabricated a PIC with an evanescent field probe (Fig. 5(a)) that consists of a bus-coupled photonic loop, where the U-shaped bus waveguide input and output ports are routed through bridge structures to a single edge of the PIC. As shown in Fig. 5(b), the full chip is a multi-function chip that includes both an electrical current trace for a magnetic trap and a bus-coupled photonic loop for an EFOT, part of continuing work to demonstrate a guided matter-wave interferometer. The bus and loop ridge waveguides are exposed to vacuum near the center of the chip by removing the $SiO_2$ cladding. A $Al_2O_3$-protected gold metal mirror is additionally deposited on the surface of the chip for a mirror MOT. Finally, as described in Sec. 2.1, an interposer chip is anodically bonded to the back of the PIC to provide UHV-compatible electrical feedthroughs for connecting the magnetic wire trap to electrical contacts on the edges of the chip.

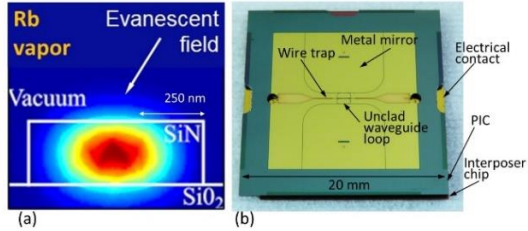

Fig. 5. Evanescent field interrogation. (a) Example simulation of evanescent field from vacuum-clad SiN ridge waveguide on SiO$_2$ substrate, showing transverse mode extending into vacuum. Colors indicate relative intensity on a linear scale. (b) Photograph of multi-function photonic chip comprising an unclad photonic ridge waveguide loop structure, an electrical current trace for magnetic trapping in proximity to the waveguide loop, and a protected mirror gold coating over most of the surface of the chip for the purpose of forming a mirror-MOT. The grey a-Si deposition required for anodic bonding is evident around the perimeter of the chip. Also shown is an interposer chip and UHV-compatible electrical feedthroughs. Note that this chip possesses a mirror image copy of the PIC that has been used to test experimental variables in the ridge waveguide design.

## 3. Atom-interrogation results

### 3.1 Grating probe experiment

For a proof-of-operation experiment of the integrated grating probe, we used the PIC-cuvette assembly shown in Fig. 1(a), where the PIC is anodically bonded to a UHV-backed, rubidium-filled cuvette via the stack-up described in Sec. 2.1. Light is edge-coupled to the air-side PIC by using a polarization maintaining, tapered optical fiber mounted in a ceramic ferrule and positioned with a precision XYZ translation stage. In Fig. 6(a) we show a top view photograph of the tapered optical fiber in very close proximity of the PIC edge for coupling light into the waveguide. The light propagates through the air/vacuum interface via the optical waveguide bridge structure described in Sec. 2.2 and out-couples as a free-space beam into the vacuum via the grating, with no glass interposer.

An $^{85}$Rb MOT is formed inside the cuvette using a six-beam conventional MOT configuration. In Fig. 6(b) we show a photograph of the MOT located in the path of the grating-emitted free-space probe beam (an inset diagram is provided for clarity). Here, the grating probe beam is on resonance with an $^{85}$Rb D2 transition and is evident via fluorescence. A photodetector is placed at the end of the probe beam to perform absorption spectroscopy. In Fig. 6(c) we display the photodetector trace, showing the interaction of the probe beam and $^{85}$Rb MOT, which is indicated by sub-Doppler hyperfine structure visible on top of a Doppler-broadened background. In Fig. 6(d) we show a similar trace with the MOT turned off, in which the sub-Doppler features disappear.

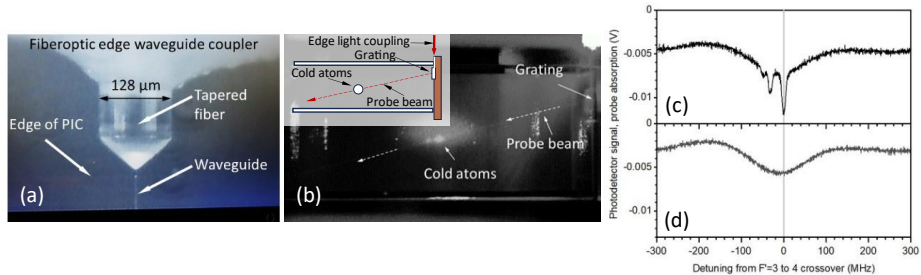

Fig. 6. Proof-of-concept experiment with grating-emitted free-space probe directed into UHV cuvette with no glass windows in between the MOT and the grating. (a) Top view photograph of PIC edge waveguide coupling using a polarization maintaining tapered optical fiber. (b) Fluorescence image of interaction of probe beam from PIC grating and a $^{85}$Rb MOT (labeled "Cold atoms"). Inset has a diagram for clarity. (c) Photodetector trace of interaction of the probe beam and MOT, showing sub-Doppler hyperfine structure, as well as a Doppler-broadened peak from thermal background vapor. (d) Photodetector trace with the MOT turned off, in which the sub-Doppler features disappear.

## 3.2 Waveguide evanescent field experiment

For a proof-of-operation experiment in one of our builds using an evanescent field probe, we demonstrate the interaction of rubidium vapor and an evanescent field from a ridge waveguide. For these experiments, we use a multi-function electronic/photonic chip that is anodically bonded to a cuvette with no glass interposer, as shown in Fig. 1(b). The cuvette is attached to a vacuum pump and filled with a natural abundance of rubidium, as described in Sec. 3.1. We have simulated the evanescent field of the ridge waveguide with the results shown in Fig. 5(a) and overall design described in Sec. 2.4. Probe light is coupled into the input waveguide using a polarization maintaining tapered optical fiber, similar to that shown in Fig. 6(a). The probe light is collected from the output waveguide using a multimode 200 µm, 0.5 NA optical fiber that is connected to an APD detector (Thorlabs APD410A). The input light is polarization-modulated, vertical to horizontal, by using an electro-optic amplitude modulator (Thorlabs EO-AM-NR-C1). We use the strong polarization dependence of the propagation through the waveguide, combined with a lock-in amplifier, to increase detection sensitivity and aid in differentiating waveguided light from stray light that may additionally couple into the collection fiber. In Fig. 7(a) we diagram how we have integrated this multi-function chip/cuvette vacuum assembly into an overall system suitable for generating a mirror MOT and other forms of magnetic trapping. In Fig. 7(b) we show a top view photograph of the PIC on the multi-function chip with added light loss in the unclad regions of the bus and loop waveguides near the center of the PIC. It is in this region that we expect an evanescent field interaction with the rubidium atoms inside the enclosure. In Fig. 7(c) we show a plot of the waveguide transmission when the probe laser is scanned across the D2 rubidium line, observing a rubidium spectrum from the interaction of the evanescent field of the waveguide with the rubidium vapor inside the cuvette, when vapor pressures are near $10^{-7}$ to $10^{-6}$ Torr. The plot also shows a signal from a reference cell and from an auxiliary free-space probe beam in the cuvette. Finally, we have generated a mirror MOT several hundred microns above the low-finesse photonic loop, shown in Fig. 7(d). These experiments are in progress to study the optical evanescent field and MOT interaction toward an application of guided matter-wave Sagnac interferometry.

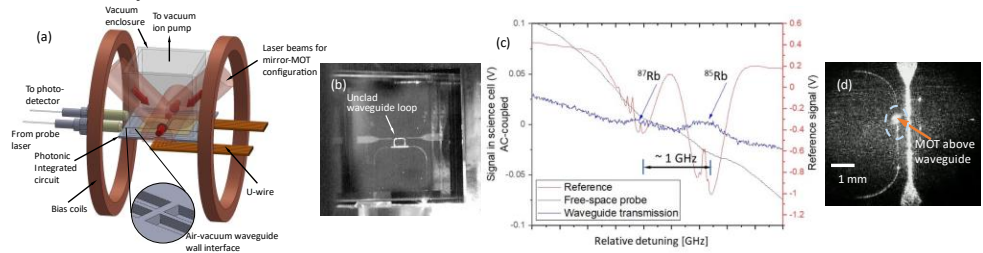

Fig. 7. Using an anodically bonded multi-function photonic/electronic chip for interaction of evanescent field from a SiN ridge waveguide with rubidium: (a) Experimental testbed consists of a multi-function photonic/electronic chip that is anodically bonded to a vacuum enclosure, forming one of its walls. The PIC includes a SiN ridge waveguide with SiN air-vacuum wall interface structures formed across the waveguide (inset) to form a UHV-hermetic seal with the vacuum enclosure. Optical fibers from the probe laser and to the photodetector are edge coupled to the PIC waveguides. The waveguide is clad with silicon dioxide but exposed to vacuum near the center of the PIC for experimenting with evanescent optical field interactions with hot and cold rubidium atoms. The PIC also includes a protected thin-film metal mirror to form a mirror MOT. (b) Top-down photograph of operational PIC, looking down on an angle through the glass cuvette. The arrow indicates edge-waveguide-coupled 780-nm light leaking from unclad areas of the U-shaped bus and loop waveguides at the center of the PIC, the region in which we expect an evanescent field to also extend into the vacuum. (c) Plot showing the interaction of the evanescent field of the waveguide with rubidium vapor inside the vacuum cuvette (blue trace). Black trace is a reference free-space probe in the same cuvette, and the red trace is a probe of a separate reference cell in a saturation-absorption configuration. Both reference probes are taken simultaneously and stem from the same probe laser as the evanescent probe, and all three signals show two prominent rubidium isotope lines. (d) Photograph of mirror MOT on top of unclad optical waveguide loop for studying evanescent field and MOT interactions. Work is in progress to close the longitudinal gap between the trapped-atom location and the PIC surface (dimension not shown).

## 4. Conclusions

We have demonstrated an approach to interfacing a photonic chip and a UHV envelope by epoxy-free anodic bonding, where the photonic chip forms one of the walls of the vacuum envelope. This approach enables delivery of light into a quantum sensor by eliminating free-space optics and optical windows and provides for a batch-fabricable UHV-compatible seal without epoxies or other sealing materials. We performed two proof-of-operation experiments: the first one by generating a free-space beam from a grating and interacting with a MOT, and the second one by interacting an evanescent field from a ridge optical waveguide with a rubidium vapor. This second photonic chip has additionally been metallized and used to generate a mirror MOT, which is an important first step for trapping atoms in an EFOT for the purpose of demonstrating a guided Sagnac matter-wave interferometer. Such wafer-compatible bonding techniques can be extended for interfacing ion traps with vacuum envelopes, as well as interfacing PICs with alkali vapor cells for applications such as quantum computing and quantum sensing.

**Funding.** This work was internally funded.

**Acknowledgments.** We would like to thank Mr. Doug Coombs for supporting the PIC postprocessing and edge polishing; Dr. Gavin Fields, Dr. Josh Frechem, Mr. Louis Sofia for supporting the experimental setups.

**Disclosures.** The authors declare no conflicts of interest.

**Data availability.** Data underlying the results presented in this paper are not publicly available at this time but may be obtained from the authors upon reasonable request.